\documentclass[10pt, conference, letterpaper]{IEEEtran}
\IEEEoverridecommandlockouts
\usepackage{siunitx}
\usepackage{ifpdf}
\usepackage{cite}
\usepackage{amsmath,amssymb,amsfonts}
\usepackage{graphicx}
\usepackage{textcomp}
\usepackage{booktabs} 
\usepackage{verbatim}
\usepackage{graphics} 
\usepackage{epsfig} 
\usepackage{tikz}
\usepackage{epstopdf}
\usepackage{subfigure}
\usepackage{balance}
\usepackage{comment}
\usepackage{booktabs}
\usepackage{amsmath}
\usepackage{amssymb}
\usepackage{amsfonts}
\usepackage{bm}
\usepackage{colortbl}
\usepackage{tabularx}
\usepackage{color}
\usepackage{xspace}
\usepackage{graphicx}    
\usepackage{url}         
\usepackage[ruled,vlined,linesnumbered,resetcount]{algorithm2e}
\usepackage{algpseudocode}
\usepackage{placeins}
\usepackage{multirow}
\usepackage{caption}
\usepackage{enumitem}
\usepackage{leading}
\usepackage{pgfplots}
\usepackage{xcolor}
\usepackage{diagbox}
\usepackage{threeparttable}
\usepackage{makecell}

\usepackage{mathtools}
\DeclareMathOperator*{\argmin}{arg\,min}

\def\BibTeX{{\rm B\kern-.05em{\sc i\kern-.025em b}\kern-.08em
    T\kern-.1667em\lower.7ex\hbox{E}\kern-.125emX}}

\newcommand{\SystemName}{\texttt{LLMKey}\xspace}

\begin{document}
\title{LLMKey: LLM-Powered Wireless Key Generation Scheme for Next-Gen IoV Systems}

\DeclareRobustCommand*{\IEEEauthorrefmark}[1]{%
  \raisebox{0pt}[0pt][0pt]{\textsuperscript{\footnotesize #1}}%
}
\author{\IEEEauthorblockN{Huanqi Yang\IEEEauthorrefmark{},
Weitao Xu\thanks{* Weitao Xu is the corresponding author.}\IEEEauthorrefmark{*}
}
\IEEEauthorblockA{
\IEEEauthorrefmark{}Department of Computer Science, City University of Hong Kong
}
\vspace{-0.3in}
}

\maketitle
\vspace{-0.3in}
\hyphenation{op-tical net-works semi-conduc-tor}
\pagestyle{empty}
\begin{abstract}
Wireless key generation holds significant promise for establishing cryptographic keys in Next-Gen Internet of Vehicles (IoV) systems. However, existing approaches often face inefficiencies and performance limitations caused by frequent channel probing and ineffective quantization. To address these challenges, this paper introduces \SystemName, a novel key generation system designed to enhance efficiency and security.
We identify excessive channel probing and suboptimal quantization as critical bottlenecks in current methods. To mitigate these issues, we propose an innovative large language model (LLM)-based channel probing technique that leverages the capabilities of LLMs to reduce probing rounds while preserving crucial channel information. Instead of conventional quantization, \SystemName adopts a perturbed compressed sensing-based key delivery mechanism, improving both robustness and security.
Extensive evaluations are conducted in four real-world scenarios, encompassing V2I (Vehicle-to-Infrastructure) and V2V (Vehicle-to-Vehicle) settings in both urban and rural environments. The results show that \SystemName achieves an average key agreement rate of 98.78\%, highlighting its effectiveness and reliability across diverse conditions.

\end{abstract}

\begin{IEEEkeywords}
Physical layer key generation, large language model, Internet of Vehicles
\end{IEEEkeywords}

\section{Introduction}
\label{intro}
\subsection{Background and Limitations}
\begin{figure}
	\centering
	\includegraphics[width=3in]{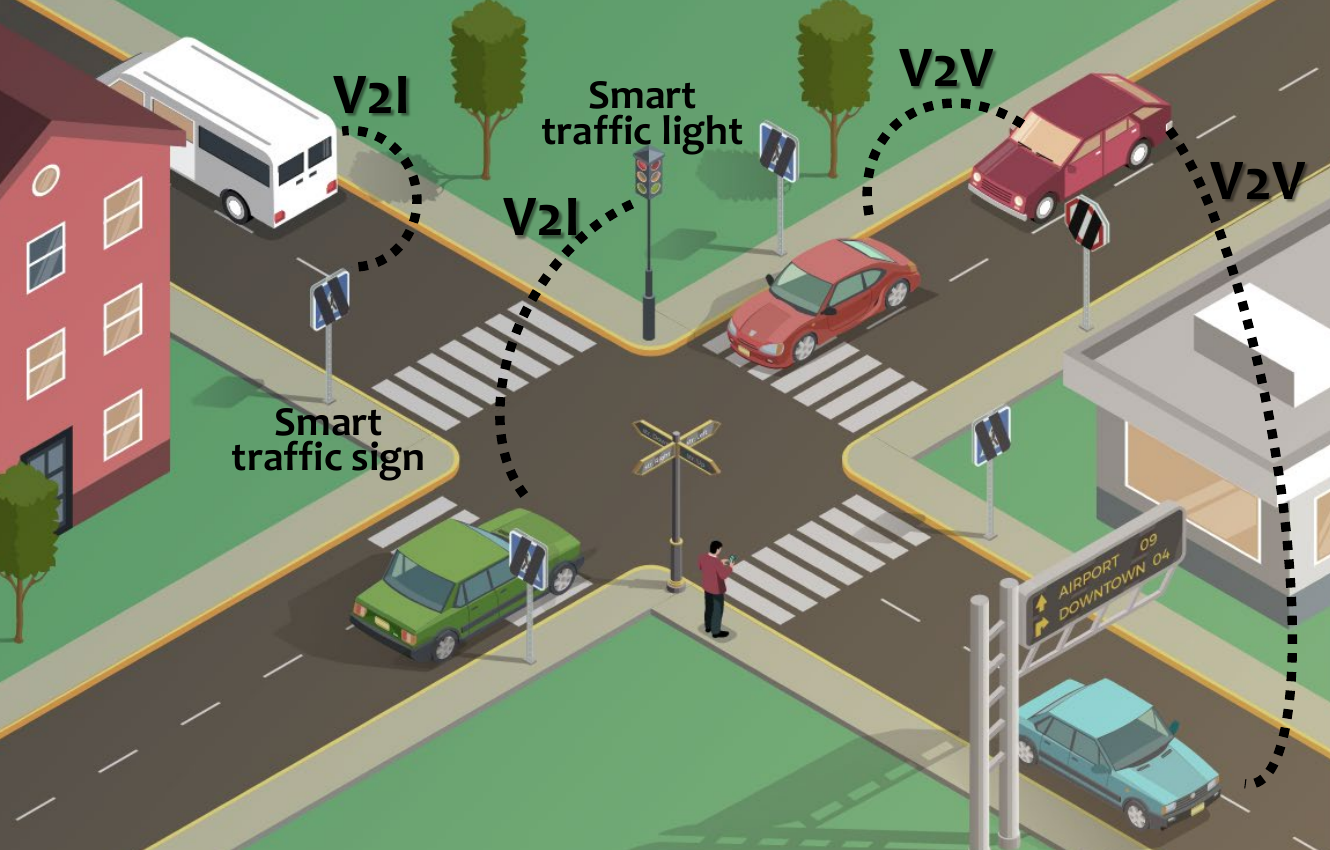}
	\caption{IoV communication scenarios.}
	\label{fig:v2x}
	\vspace{-0.25in}
\end{figure}

The Internet of Vehicles (IoV) has rapidly evolved to facilitate seamless connectivity between various entities in vehicle networks, aiming to enhance urban transportation, minimize accidents, and bolster traffic surveillance. This technology's pervasive nature has drawn significant interest. As shown in Fig.~\ref{fig:v2x}, IoV integrates diverse objects using wireless communication methods, supporting multiple interaction models such as vehicle-to-vehicle (V2V) and vehicle-to-infrastructure (V2I).

Aforementioned Vehicle-to-everything (V2X) frameworks necessitate the frequent exchange of critical real-time data, including speed and location, across vehicles and essential infrastructure components over wireless channels. Safeguarding this data transmission is paramount to maintaining the integrity of vehicular systems and ensuring passenger safety. For instance, adversaries can intercept communications and compromise control units by decoding passcodes, potentially causing catastrophic outcomes or fatalities~\cite{hasrouny2017vanet}. Consequently, the need for robust key establishment processes has become important, allowing two authenticated nodes to create a secure key over public channels. However, conventional approaches like public key infrastructure (PKI) and pre-shared key (PSK) are inadequate for IoV networks due to two primary factors: 1) the transient and highly dynamic nature of IoV systems makes sustained PKI availability impractical; 2) PSK methods are rigid, difficult to scale, and susceptible to theft by malicious actors. The inherently mobile and decentralized characteristics of IoV present significant challenges for secure key generation and management.

To address the limitations discussed earlier, physical layer key generation leveraging channel reciprocity has gained traction as a viable method for establishing secure cryptographic keys between two IoV communication nodes. This approach is based on the concept that certain channel properties, such as the Received Signal Strength Indicator (RSSI), will exhibit similarity at both ends if data packets are exchanged within the channel's coherence time~\footnote{Coherence time refers to the period during which the channel remains relatively stable}~\cite{jana2009effectiveness}.
However, the necessity for long-distance communication restricts the applicability of many short-range wireless technologies (e.g., WiFi, Zigbee) in generating secret keys for IoV environments. LoRa, a widely-used low-power wide-area network (LPWAN) technology, offers a compelling alternative by enabling long-range communication suitable for IoV key generation. Despite recent advancements in this area~\cite{junejo2021lora,ruotsalainen2019experimental,xulorakey,yang2022vehicle,jiayaoipsn}, our investigation highlights two critical and unresolved challenges:

\textbf{(1) Excessive channel probing process.}
Physical-layer key generation relies heavily on repeated channel probing, requiring both legitimate parties to exchange numerous probing packets over multiple rounds to measure the channel characteristics based on the principle of channel reciprocity. To derive a 128-bit cryptographic key, the nodes must transmit a sufficient volume of effective packets. However, this process is not only energy-intensive for communication nodes but also increases the risk of eavesdropping by malicious actors. As a result, the current channel probing methods fail to deliver both energy efficiency and robust security for physical-layer key generation in IoV networks.

\textbf{(2) Inefficient quantization process.} 
Once channel measurements are collected, the subsequent step is quantization, which translates these measurements into binary values. Despite the development of various quantization techniques~\cite{junejo2021lora,ruotsalainen2019experimental,xulorakey,yang2022vehicle,jiayaoipsn}, the process remains inherently lossy and susceptible to errors~\cite{xi2016instant}. As a result, key generation systems relying on quantization may require a higher volume of packet exchanges to gather sufficient channel data, often necessitating an additional reconciliation phase to correct errors. This, in turn, can compromise system efficiency and diminish overall robustness.

\subsection{Contributions}
In this paper, we design and implement a novel physical layer key generation system for IoV networks named \SystemName, which essentially addresses the above challenges.

To achieve energy-efficient and secure channel probing, we propose an large language model (LLM)-enabled skipped channel probing and channel information recovery method. Drawing inspiration from the increasing potential of LLMs to enhance data management and operational efficiency in smart transportation systems~\cite{yang2024transcompressor}, our goal is to explore the inferential and reconstructive capabilities of LLMs. By providing skipped channel measurement data and a simple prompt to the cloud, the LLM processes this input to generate accurate and efficient reconstructions of the original channel measurements. This approach minimizes the need for excessive probing, reducing energy consumption and enhancing security in IoV networks.

To tackle the second challenge, we adopt a Perturbed Compressed Sensing (PCS)-based key delivery technique, recently introduced as an efficient way to transmit secret keys between devices~\cite{yang2023chirpkey}. This approach utilizes PCS to facilitate key delivery by accommodating slight discrepancies between the matrices of the sender and receiver. We leverage channel measurements—similar yet not identical data—to construct these matrices, where one matrix compresses the secret key and the other decompresses it.
By eliminating the need for quantization and information reconciliation, this method enhances both system efficiency and robustness in secret bit agreement. Unlike traditional quantization-based approaches, the secret key is generated through a random key generator, allowing for the creation of highly random and secure keys.

We conduct extensive evaluations in both V2I and V2V settings in both urban and rural environments. Results show that \SystemName achieves a high matching rate. 
In summary, this paper makes the following contributions.
    
\begin{itemize}
    \item We propose the first LLM-powered physical-layer key generation scheme, enhancing the accuracy, security, and efficiency of secret key generation in IoV systems.
    \item \SystemName demonstrates that LLMs can act as zero-shot reconstructors of channel measurement data, effectively interpreting and recovering information without the need for fine-tuning or domain-specific training.  
    \item We conduct extensive experiments to evaluate \SystemName in different real environments. Comprehensive evaluations show that our system can achieves an average key agreement rate of 98.78\%.
\end{itemize}


\section{Related Work}
\label{sec:Background}
\subsection{Large language models}
Recent breakthroughs in large language models (LLMs) have significantly propelled advancements in artificial intelligence and natural language processing. Models like OpenAI's GPT-3 and Meta's LLaMA 2~\cite{touvron2023llama} have demonstrated remarkable proficiency in producing contextually appropriate and coherent text, excelling in tasks such as multilingual translation, question answering, and code generation.
The evolution of neural language models (NLMs)\cite{arisoy2012deep}, alongside earlier LLMs like GPT-2\cite{radford2019language} and BERT~\cite{devlin2018bert}, laid the groundwork for these advancements. Over time, language models have grown in sophistication and capability, as seen with more recent models like PaLM~\cite{chowdhery2023palm} and GPT-4~\cite{achiam2023gpt}.

\subsection{Wireless Key Generation} 

Wireless key generation~\cite{zhang2016key} has attracted significant interest over the past few decades. Numerous systems have been developed across various wireless technologies, including Wi-Fi~\cite{xi2016instant, liu2013fast}, Zigbee~\cite{Aonowireless2005}, LoRa~\cite{zhang2018channel, ruotsalainen2019experimental, xulorakey, jiayaoipsn, yang2022vehicle}, and Bluetooth~\cite{premnath2014secret}. Researchers have explored diverse physical-layer attributes, such as Channel State Information (CSI)\cite{xi2016instant, liu2013fast}, Received Signal Strength Indicator (RSSI)\cite{yang2022vehicle, jiayaoipsn}, and phase~\cite{wang2011fast}.
For instance, TDS~\cite{xi2016instant} utilized Wi-Fi CSI to derive cryptographic keys for mobile devices by leveraging channel characteristics. To further improve CSI reciprocity, Liu et al.~\cite{liu2013fast} employed channel responses from multiple Orthogonal Frequency-Division Multiplexing (OFDM) subcarriers. This approach was complemented by a Channel Gain Complement (CGC) scheme to facilitate more reliable key generation.

\subsection{LLMs for Time Series Analysis}
The application of LLMs to time series tasks has garnered increasing attention from researchers. For instance, TIME-LLM repurposed an existing LLM for time series forecasting while maintaining the original architecture of the language model~\cite{jin2023time}. 
Similarly, LLMTIME~\cite{gruver2024large} leveraged pretrained LLMs for continuous time series prediction by encoding numerical values as text and generating forecasts through text-based completions. However, this method primarily focuses on the temporal dimension, overlooking spatial dependencies in the data.
To address spatial considerations, GATGPT integrated a graph attention network with GPT to perform spatio-temporal imputation~\cite{chen2023gatgpt}. 
Furthermore, researchers have explored the use of LLMs for interpreting sensor data to facilitate real-world sensing~\cite{ji2024hargpt,yang2024you,yang2024transcompressor}.

\begin{figure*}
    \centering
    \includegraphics[width=5.5in]{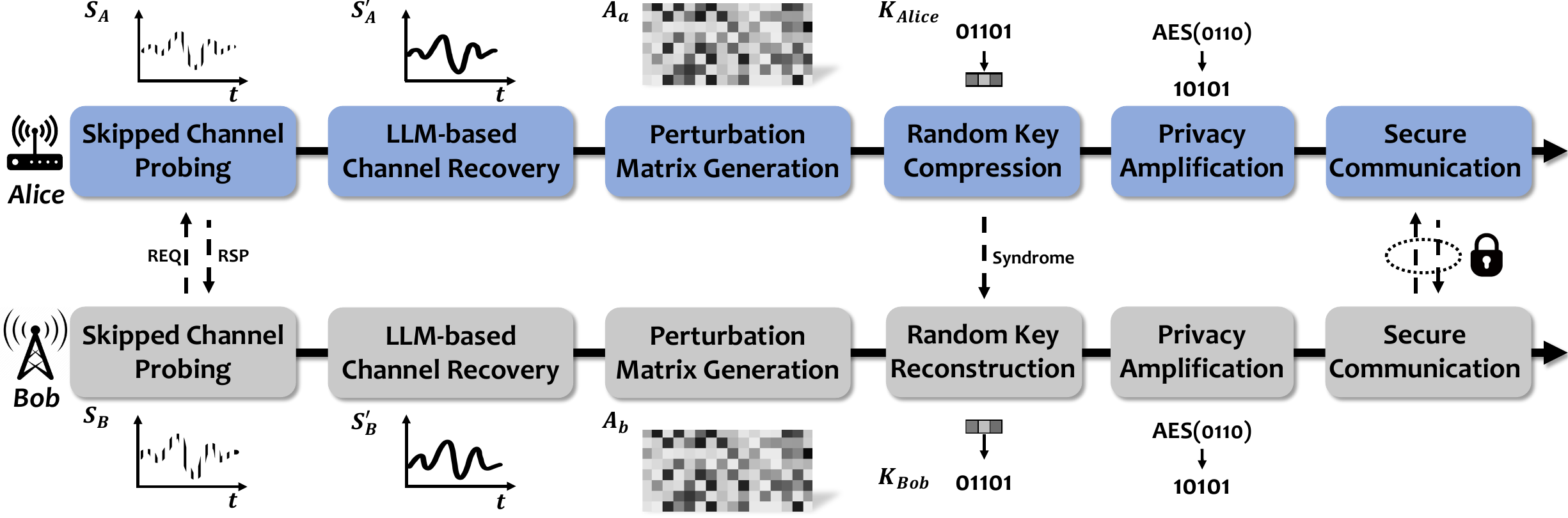}
    \caption{System workflow.}
    \label{fig:workflow}
    \vspace{-0.3in}
\end{figure*}

\section{System Model}
\label{sec:model}
In an IoV network, we consider two devices, Alice and Bob, aiming to establish a shared key to secure their communication. Both devices are equipped with LoRa communication modules, and no pre-shared secrets exist between them. They follow the process outlined in Fig.~\ref{fig:workflow} to generate cryptographic keys step by step. This process involves skipped channel probing, LLM-based channel recovery, perturbed measurement matrix generation, key compression and reconstruction, and privacy amplification. Notably, the LLM-based recovery and key reconstruction are performed using a server-based API service, which is a reasonable assumption due to the server's enhanced computational capabilities and robust security protocols.
Once the key is successfully generated, it is applied to encrypt and decrypt data, ensuring secure communication between the two devices.




%

\section{System Design}
\label{sec:signal}
\subsection{Skipped Channel Probing}
Channel probing is a critical process for measuring channel characteristics, which forms the foundation for key generation in wireless communication systems. In \SystemName, we utilize adjacent rRSSI (arRSSI)~\cite{yang2022vehicle} as the channel indicator. Skipped channel probing involves randomly performing probing for only a specified fraction \( \alpha \) of the total potential probing instances. For example, if \( \alpha = 0.1 \), approximately 10\% of the total possible probing instances are executed, while the rest are skipped randomly.
The number of probing instances performed, \( n_{\text{probed}} \), for sequences \( S_{A} \) and \( S_{B} \) can be calculated as:
\[
n_{\text{probed}} = \lfloor \alpha \cdot n_{\text{total}} \rfloor,
\]
where \( n_{\text{total}} \) represents the total number of potential probing instances, and \( \lfloor \cdot \rfloor \) denotes the floor function, rounding down to the nearest whole number.
Skip channel probing effectively reduces the overhead of frequent probing, alleviating the load on communication channels and conserving energy. This technique enhances efficiency, particularly in resource-constrained environments and large-scale deployments, by maintaining sufficient channel information without excessive probing.


\subsection{Data Rescaling}
After the selectively probed channel data is collected, it undergoes rescaling to prepare for further processing and analysis. The rescaling process ensures the data conforms to a standard scale, enhancing consistency and compatibility with subsequent operations. Specifically, the channel data sequences \( S_{A} \) and \( S_{B} \) are rescaled to the range \([0, 1]\).  
The rescaling of a data value in \( S_{A} \) or \( S_{B} \), originally in the range \([S_{\text{min}}, S_{\text{max}}]\), is performed using the following formula:  
\[
S_{\text{scaled}} = \frac{S - S_{\text{min}}}{S_{\text{max}} - S_{\text{min}}},
\]
To optimize the data for lightweight processing by LLMs, the rescaled values are truncated to retain only two decimal places. This reduces the token length, enhancing efficiency and reducing computational overhead. The truncation process is defined as:  
\[
S_{A\text{truncated}} = \left\lfloor 100 \cdot S_{A\text{scaled}} \right\rfloor / 100,
\]
\[
S_{B\text{truncated}} = \left\lfloor 100 \cdot S_{B\text{scaled}} \right\rfloor / 100,
\]
where \( \left\lfloor \cdot \right\rfloor \) represents the floor function, ensuring the value is truncated at two decimal places.  
The truncated and rescaled data sequences \( S_{A\text{truncated}} \) and \( S_{B\text{truncated}} \) are subsequently ready for input into LLMs, preserving essential information while maintaining efficiency in resource-constrained environments.

\subsection{LLM-based Channel Information Recovery}
\begin{figure}
    \centering
    \includegraphics[width=0.83\linewidth]{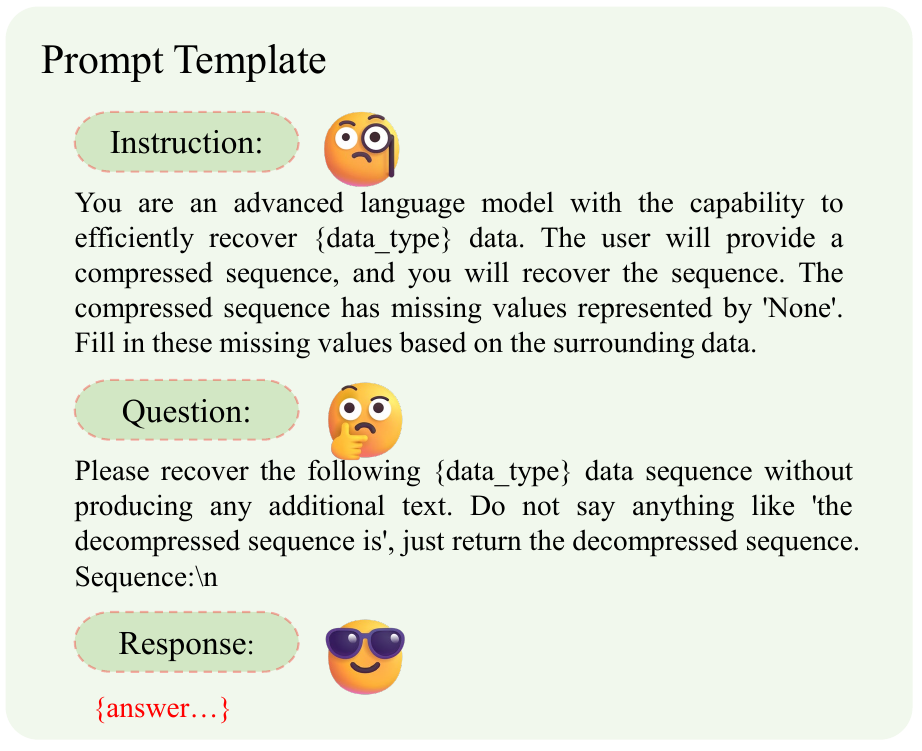}
    \caption{Chain-of-thought prompt design for \SystemName.}
    \label{fig:prompt}
    \vspace{-0.2in}
\end{figure}

To recover the skipped channel measurements, we employ an LLM to reconstruct the arRSSI data for Alice and Bob.
The relationship between the input data and the LLM's output can be mathematically expressed as follows:
\[
S_{A}^{\prime} = \text{LLM}(S_{A\text{truncated}}), S_{B}^{\prime} = \text{LLM}(S_{B\text{truncated}}),
\]
where \( S_{A\text{truncated}} \) and \( S_{B\text{truncated}} \) represents the truncated version of the data, as described in the previous sections. The function \( \text{LLM} \) encapsulates the processing capabilities of the LLM, which computes the output \( S_{A}^{\prime} \) and \( S_{B}^{\prime} \) based on the input data.
The structure of our prompt, as depicted in Fig.~\ref{fig:prompt}, follows a clean and concise design. It begins with a simple instruction that directs the LLM to draw from its knowledge of channel measurement data. Next, a question is presented, offering detailed context about the data collection process, the sequence of measurements, and the specific type of channel information involved.
The prompt concludes with a clear directive: "Do not say anything like 'the decompressed sequence is', just return the decompressed sequence." This deliberate phrasing ensures that the output remains streamlined and easy to interpret, minimizing unnecessary text and enhancing clarity for seamless analysis.

\begin{figure*}
\centering
\subfigure[$\alpha$ = 0.5, V2I.]{
\begin{minipage}[t]{0.32\linewidth}
\centering
\includegraphics[width=\linewidth]{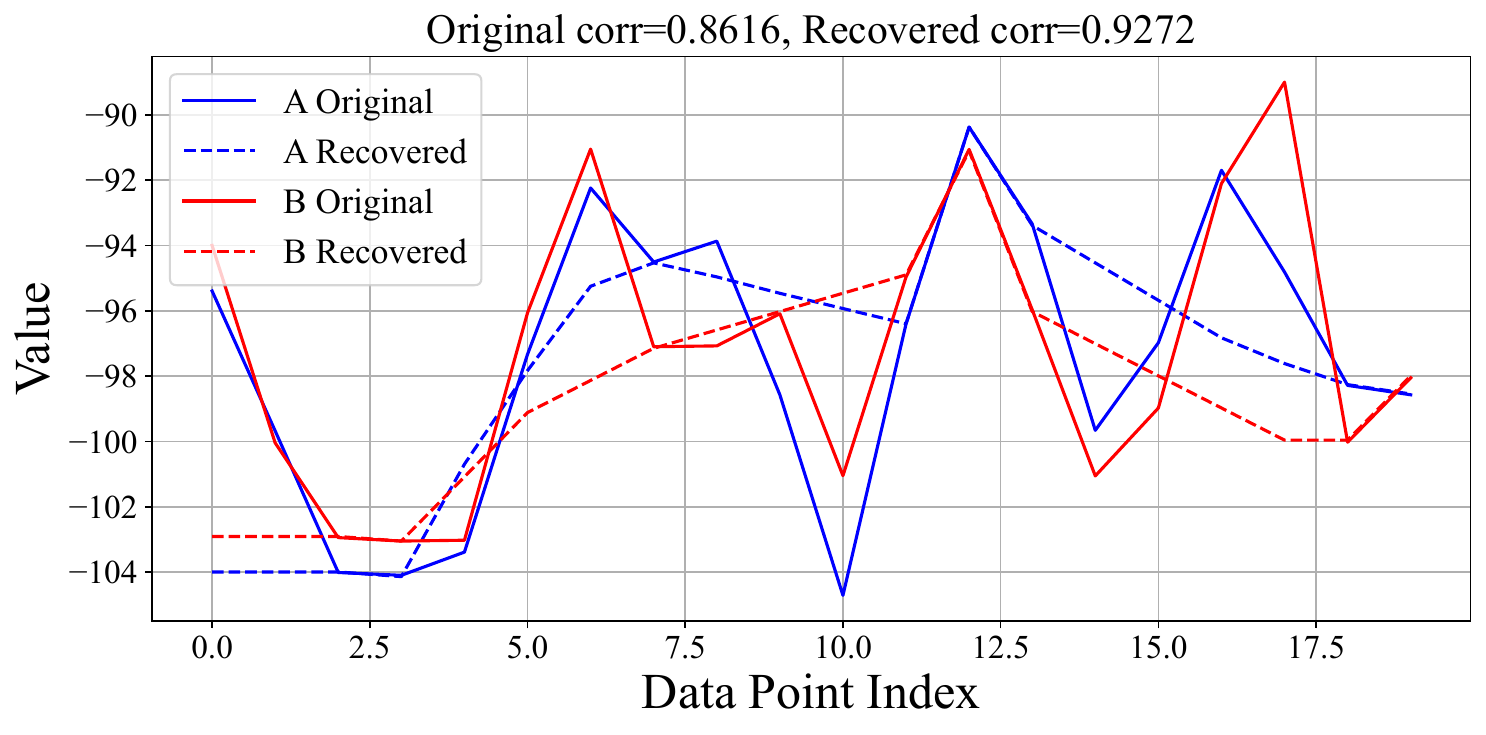}
\label{fig:v2i1}
\end{minipage}%
}%
\subfigure[$\alpha$ = 0.7, V2I.]{
\begin{minipage}[t]{0.32\linewidth}
\centering
\includegraphics[width=\linewidth]{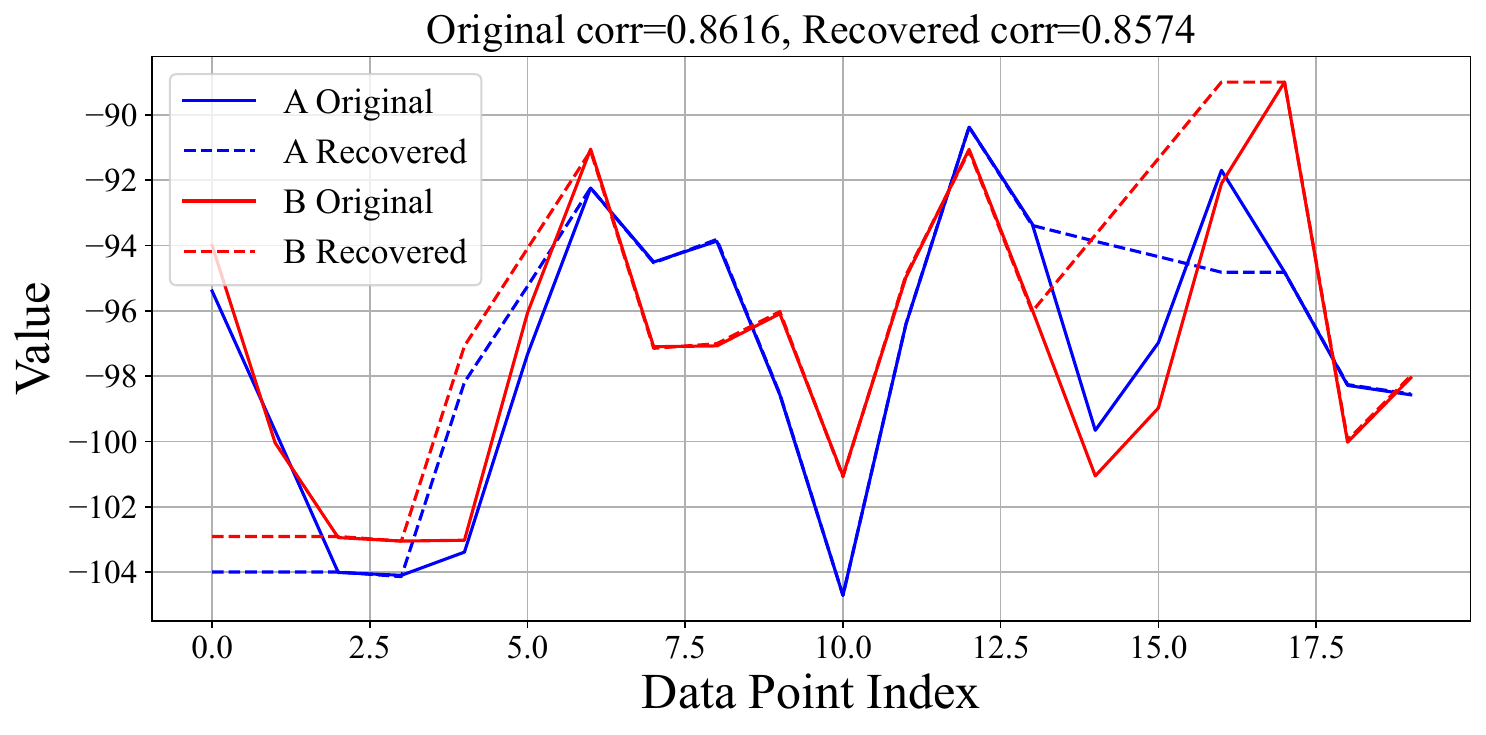}
\label{fig:v2i2}
\end{minipage}%
}%
\subfigure[$\alpha$ = 0.9, V2I.]{
\begin{minipage}[t]{0.32\linewidth}
\centering
\includegraphics[width=\linewidth]{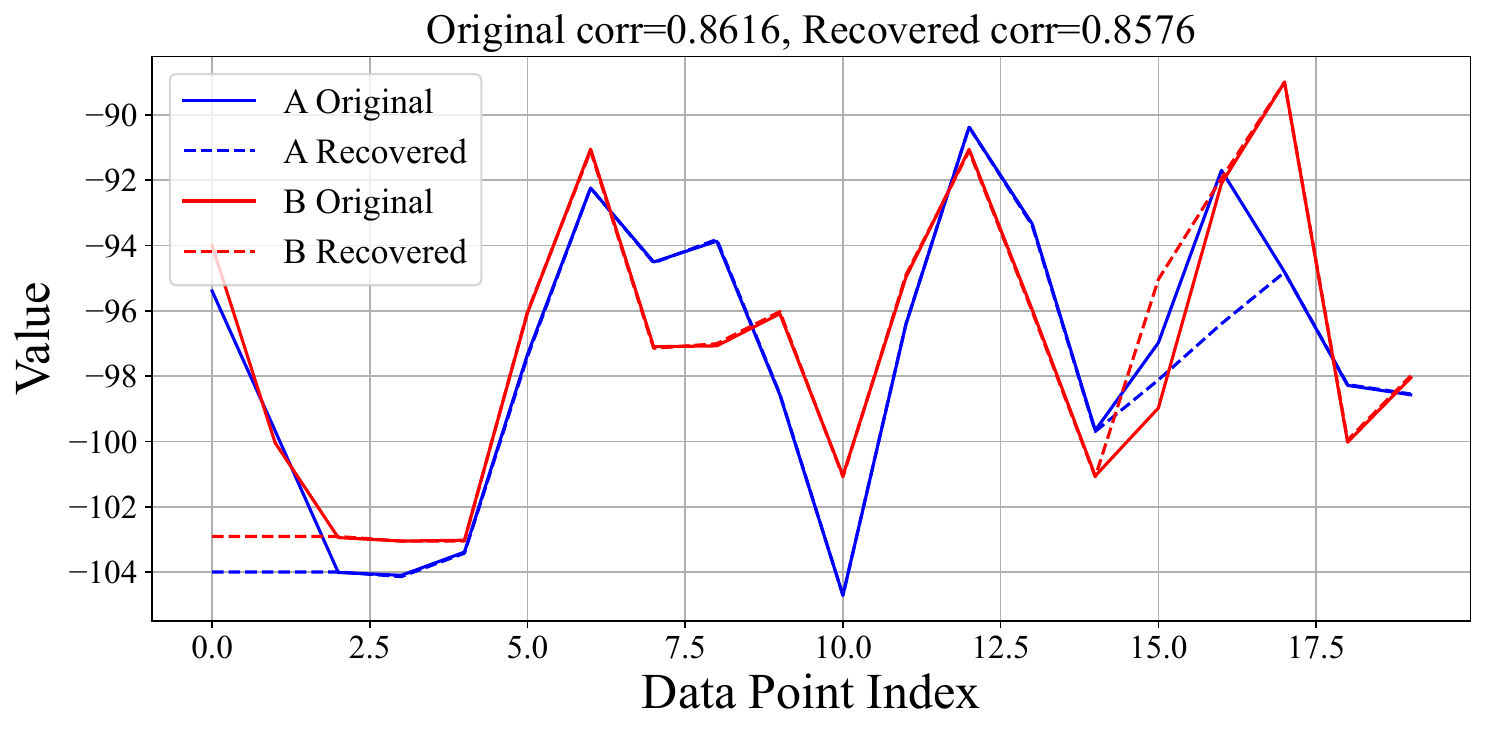}
\label{fig:v2i3}
\end{minipage}%
}%
\\
\subfigure[$\alpha$ = 0.5, V2V.]{
\begin{minipage}[t]{0.32\linewidth}
\centering
\includegraphics[width=\linewidth]{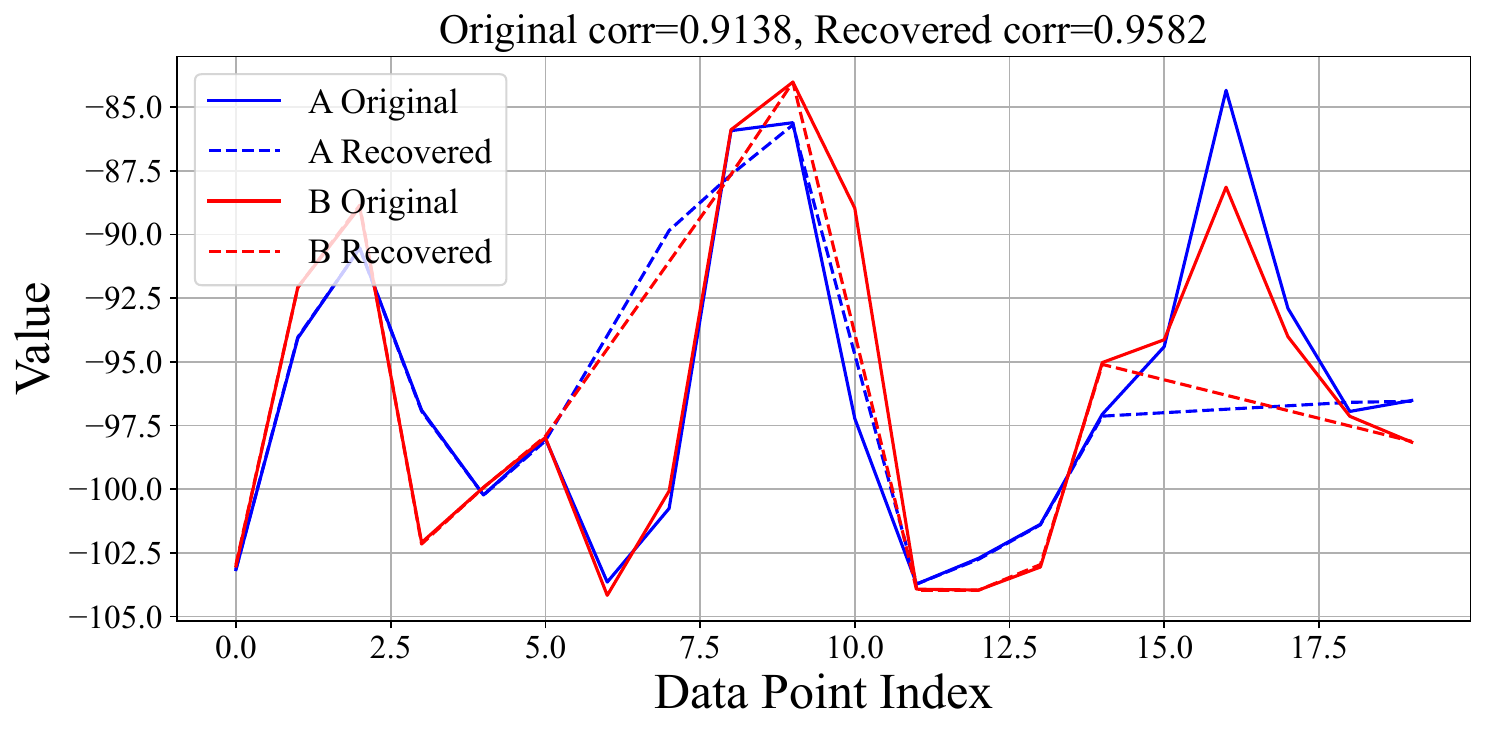}
\label{fig:v2v1}
\end{minipage}%
}%
\subfigure[$\alpha$ = 0.7, V2V.]{
\begin{minipage}[t]{0.32\linewidth}
\centering
\includegraphics[width=\linewidth]{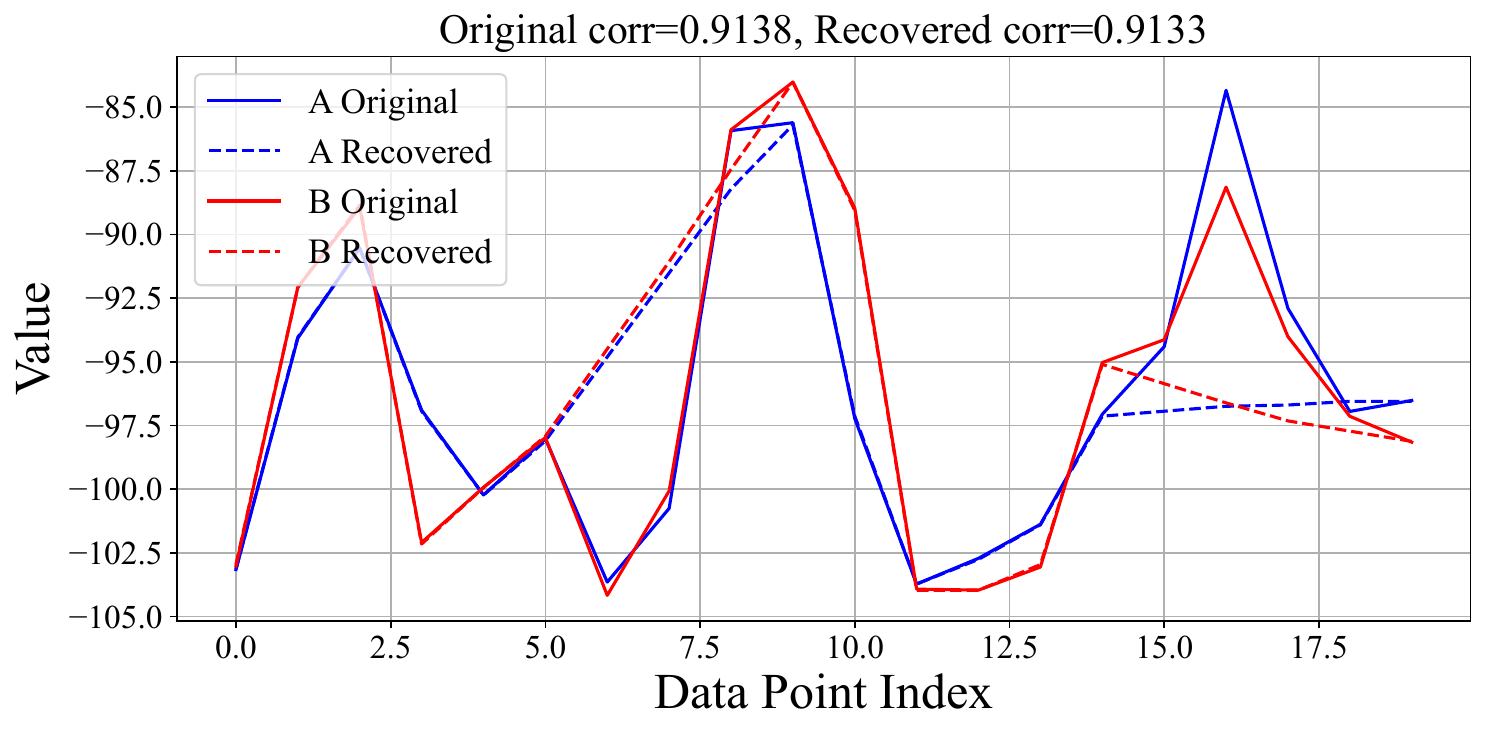}
\label{fig:v2v2}
\end{minipage}%
}%
\subfigure[$\alpha$ = 0.9, V2V.]{
\begin{minipage}[t]{0.32\linewidth}
\centering
\includegraphics[width=\linewidth]{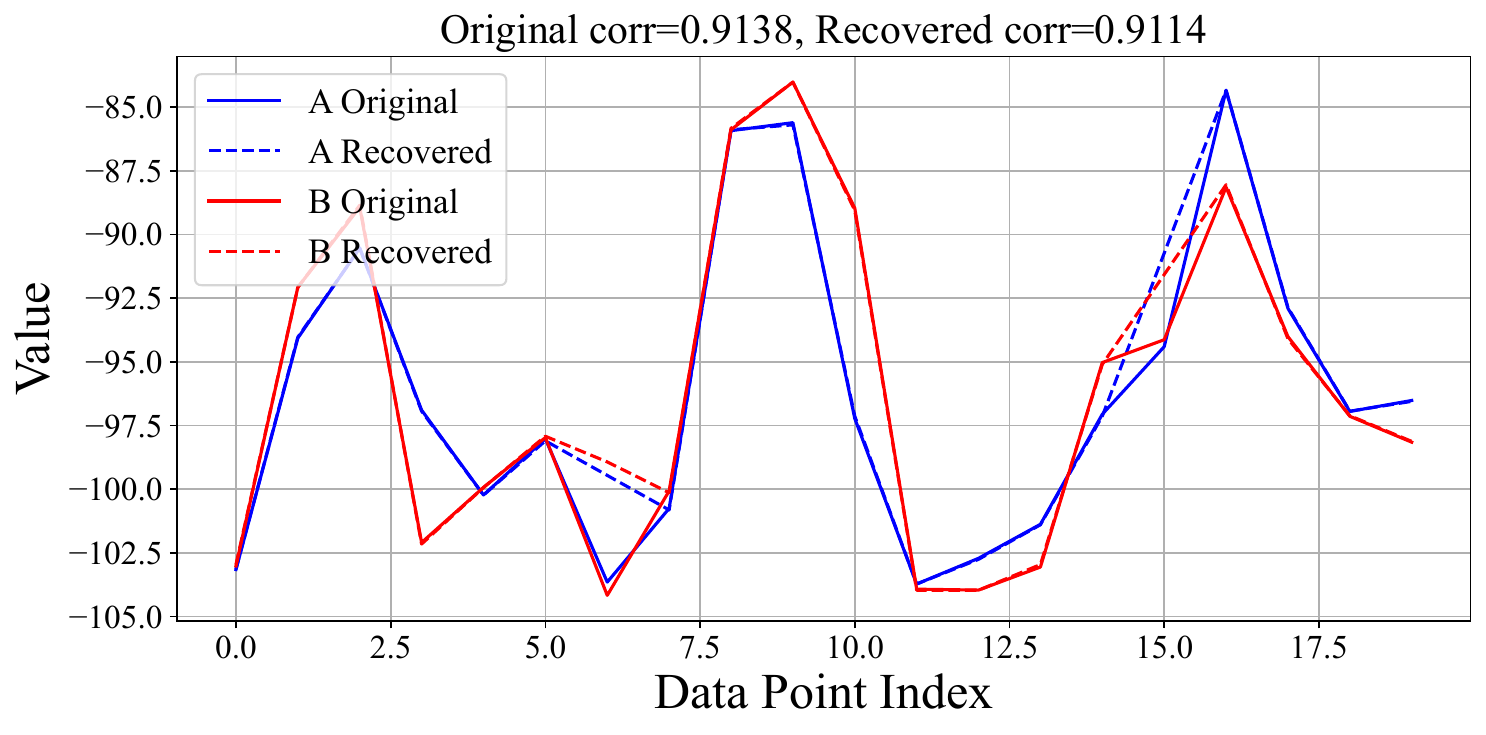}
\label{fig:v2v3}
\end{minipage}%
}%
\centering
\vspace{-0.1in}
\caption{Channel recovery results.}
\label{fig:reconstruct}
\vspace{-0.2in}
\end{figure*}

Fig.~\ref{fig:reconstruct} showcases the reconstruction results achieved using the LLM-based recovery method. The figure demonstrates that as the probing ratio $\alpha$ increases from 0.5 to 0.9, the accuracy of channel measurement data reconstruction significantly improves. This improvement can be attributed to higher probing ratios, which preserve more original data and reduce information loss during skipped probing, resulting in more precise reconstructions. Furthermore, the correlation between Alice and Bob post-recovery remains comparable to that of the original data.

\subsection{PCS-based Key Delivery} \label{sec:key}

As previously mentioned, conventional quantization is inherently lossy and can result in discrepancies between legitimate nodes~\cite{xi2016instant}. 
For example, fixed thresholds may produce differing quantized bits for Alice and Bob~\cite{jana2009effectiveness, weitaoInaudibelKey, wei20222}.
To mitigate this issue, we adopt a PCS-based key delivery approach~\cite{yang2023chirpkey}, which involves generating a measurement matrix, followed by key compression and subsequent reconstruction.

\textbf{Perturbed measurement matrix generation.}
In this work, we adopt the circulant matrix as the structured perturbation matrix due to its simplicity and ease of implementation in low-power LoRa end nodes.  
The construction process of a typical circulant matrix \( f(R) \) is as follows. First, a random vector \( R \) is generated, expressed as  
$R=\left(r_{0}, r_{1}, \cdots, r_{N-1}\right) \in R^{N}$
This vector undergoes cyclic shifts \( M \) times to form the remaining \( M-1 \) row vectors. The final circulant matrix \( f(R) \) is constructed by arranging all row vectors in a cyclically shifted manner, which is described as
\begin{equation}
\small
f(R)= \left(\begin{array}{cccc}r_{0} & r_{N-1} & \cdots & r_{1} \\ r_{1} & r_{0} & \cdots & r_{2} \\ \vdots & \vdots & \ddots & \vdots \\ r_{M-1} & r_{M-2} & \cdots & r_{M}\end{array}\right).
\end{equation}

We assume the existence of a default sensing matrix \( A_{0} \) (in \SystemName, a random Gaussian matrix is used). The objective for Alice and Bob is to utilize their channel measurements to generate a perturbation matrix that modifies this default matrix. 
However, we observe that directly applying the perturbation matrix to \( A_{0} \) can reduce reconstruction accuracy, particularly when the perturbation matrix has a high average value.  
To mitigate this issue and regulate the magnitude of the perturbation matrix, we introduce a scaling factor \( \theta \). This factor is applied to the recovered arRSSI values from Alice \( S^{\prime}_{{A}} \) and Bob \( S^{\prime}_{{B}} \), resulting in scaled perturbation matrices:  
\begin{equation}
\small
\hat{S}_{{A}} = \frac{1}{\theta} S^{\prime}_{{A}} \quad \text{and} \quad \hat{S}_{{B}} = \frac{1}{\theta} S^{\prime}_{{B}}
\end{equation}

The perturbed measurement matrices for Alice and Bob are subsequently derived by applying the generated perturbation matrix to the default matrix \( A_{0} \). This process can be expressed as:  
\begin{equation}
\small
A_{a} = A_0 (I + E_a) \quad \text{and} \quad A_{b} = A_0 (I + E_b),
\end{equation}
where \( I \) represents the identity matrix, and \( E_a = f\left(\hat{S}_{{A}}\right) \) and \( E_b = f\left(\hat{S}_{{B}}\right) \) correspond to the circulant matrices generated from the scaled arRSSI values of Alice and Bob, respectively.

\textbf{Compression by Alice.} Alice begins by generating a random binary sequence \( K_{A} \in R^{N} \). To meet the sparsity requirement of compressed sensing, she increases the sparsity of \( K_{A} \) by inserting four zero bits between each bit, producing a sparse vector \( K_{A}^{s} \).  
Alice then computes a syndrome, defined as  
\begin{equation}
\small
Syn = [Syn_{1}, Syn_{2}] = [A_{a} K_{A}^{s}, A_{a} K_{A}]
\end{equation}
The first component \( A_{a} K_{A}^{s} \) represents the compressed form of the secret key \( K_{A}^{s} \), while the second component \( A_{a} K_{A} \) serves as a mechanism for error detection and correction. Finally, Alice transmits the syndrome \( Syn \) to Bob over a public channel.

\textbf{Reconstruction by Bob.} Let \( Syn^{\prime} \) represent the received syndrome, which includes some noise, expressed as \( Syn^{\prime} = Syn + e = [Syn_{1} + e, Syn_{2} + e] \). Upon receiving the syndrome, Bob leverages the first part \( (Syn_{1} + e) \) to reconstruct a sparse key \( K_{B}^{s} \) by solving the following \( \ell_{1} \)-regularized total least squares problem:  
\begin{equation}
\label{equa:4}
\small
    \begin{gathered}
	\argmin_{{{e}, {E_p}, K_{B}^{s}}} \ \ \|{e}\|_{2}^{2}+\|{E_p}\|_{F}+\lambda\|{K_{B}^{s}}\|_{1}\\
	\text{subject to} \ \ ({A_b}+{E_p}) {K_{B}^{s}}=Syn_{1}+e,
	\end{gathered}
\end{equation}
where \( E_p \) represents the perturbation difference between \( A_b \) and \( A_a \), defined as \( E_p = A_b - A_a \). Once Bob reconstructs the sparse key \( K_{B}^{s} \), he applies a de-sparsification process to retrieve \( K_{B}^{\prime} \). Through this method, Bob derives the estimated key \( K_{B}^{\prime} \) from the first part of the received syndrome.

\begin{figure}[tbp!]
    \centering
    \includegraphics[width=0.8\linewidth]{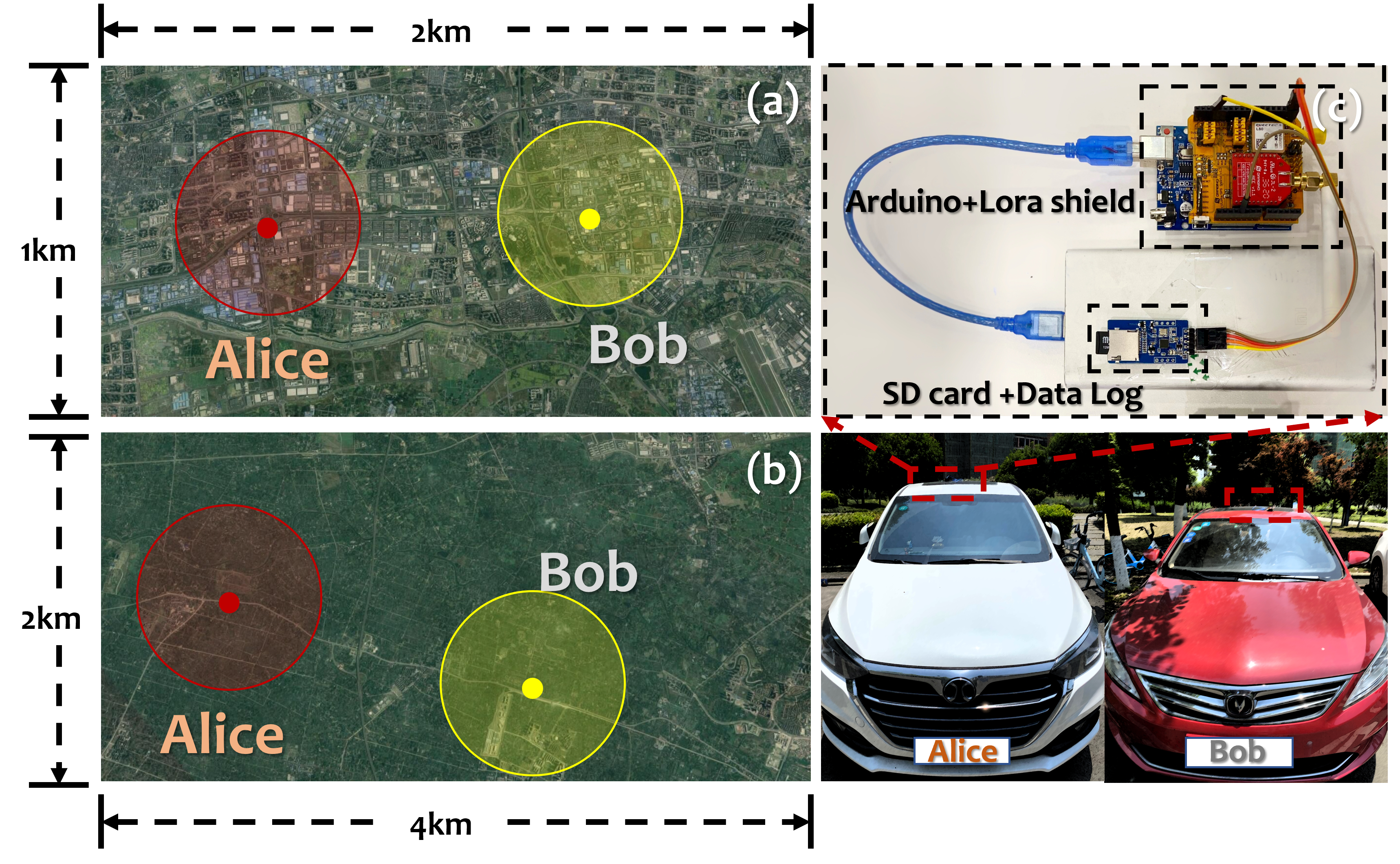}
    \vspace{-0.1in}
	\caption{Experimental setup.}
    \vspace{-0.2in}
	\label{fig:real_world}
\end{figure}

In practice, Bob’s reconstructed key \( K_{B}^{\prime} \) typically differs slightly from Alice’s original key \( K_{A} \) due to the presence of noise. To address these discrepancies, Bob leverages the second part of the syndrome to perform error correction. He begins by computing a mismatch vector \( \Delta \) as follows:  
\begin{equation}
\small
\label{equa:7}
  \begin{aligned}
    \Delta &=A_b K_{B}^{\prime}-Syn_{2} = (A_b K_{B}^{\prime}-A_a K_{A})+e \\
         &= A_b K_{B}^{\prime}-(A_b+E_p) K_A = A_b (K_{B}^{\prime}-K_A)+e \\
         &=A_b \Delta_{AB}+e,
  \end{aligned}
\end{equation}
where \( \Delta_{AB} \) represents the difference between Alice’s original key \( K_{A} \) and Bob’s reconstructed key \( K_B^{\prime} \). Since Alice and Bob are legitimate parties, the discrepancy between their keys is minimal, resulting in a sparse \( \Delta_{AB} \). Bob can recover these mismatches by solving the following \( \ell_{1} \)-regularized total least squares problem:  
\begin{equation}
\small
\label{equa:nn}
	\argmin_{{\Delta}, {e}} \|{e}\|_{2}^{2}+\lambda\|\Delta\|_{1}
	\quad \text{ subject to } {A_b} {\Delta_{AB}} = \Delta + e,
\end{equation}
This process ensures that Bob can accurately reconstruct any remaining errors, thereby aligning his key more closely with Alice’s. Using \( \Delta_{AB} \), Bob can derive \( K_{A} \) by performing the XOR operation, expressed as \( K_{B} = K_{B}^{\prime} \oplus \Delta_{AB} \), where \( \oplus \) denotes the XOR function. As a result, Alice and Bob ultimately converge on the same key, ensuring \( K_{A} = K_{B} \).


\subsection{Privacy Amplification}
Although the PCS-based key delivery method demonstrates high reliability, it inadvertently exposes some information to potential attackers since Alice and Bob must transmit the syndrome over a public channel. This issue can be mitigated using privacy amplification techniques, such as hash functions~\cite{weitaoInaudibelKey,xulorakey}.  
In \SystemName, we apply the widely adopted SHA-128 hash function to enhance the randomness and security of the final keys. Once Alice and Bob successfully generate the same key, they can utilize AES-128 or other symmetric encryption algorithms to secure their communications through encryption and decryption.

\section{Evaluation}
\label{sec:eva}
\subsection{Experimental Setup}
\label{subsec:setup}
Fig.~\ref{fig:real_world} illustrates the experimental setup and data collection process. An Arduino Uno equipped with a Dragino LoRa Shield\footnote{https://www.dragino.com/products/lora/item/102-lora-shield.html} (AVR ATmega328P, SX1278) is utilized as the communication node in the IoV environment. The experiments are conducted across four distinct IoV scenarios: V2I-Urban, V2I-Rural, V2V-Urban, and V2V-Rural.
In each scenario, two identical devices represent Alice and Bob. For V2V experiments, the devices are mounted on the rooftops of vehicles, with Alice and Bob moving randomly, resulting in distances ranging from several hundred meters to a few kilometers. In V2I scenarios, Bob is stationed on the roof of a building to emulate an infrastructure node, while Alice moves freely.
To analyze the datasets collected under different probing rates and IoV scenarios, we utilize GPT-4~\cite{achiam2023gpt}, the most advanced and capable LLM currently available. 

\textbf{Metric.} For evaluating the accuracy of recovered channel measurement data, we employ the Mean Squared Error (MSE):
\[
\text{MSE} = \frac{1}{n} \sum_{i=1}^{n} (y_i - \hat{y}_i)^2.
\]
Here, \( n \) denotes the number of data points, \( y_i \) the actual values, and \( \hat{y}_i \) the predicted values. 
For evaluating the accuracy of the generated key, we use the key agreement rate (KAR):
\[
\text{KAR} = \frac{\text{Number of matching key bits}}{\text{Total number of key bits}}.
\]
The key agreement rate measures the proportion of bits that match between the keys generated by Alice and Bob.

\subsection{Overall Performance}
\begin{figure}
\centering
\subfigure[Recovery accuracy.]{
\begin{minipage}[t]{0.48\linewidth}
\centering
\includegraphics[width=1\linewidth]{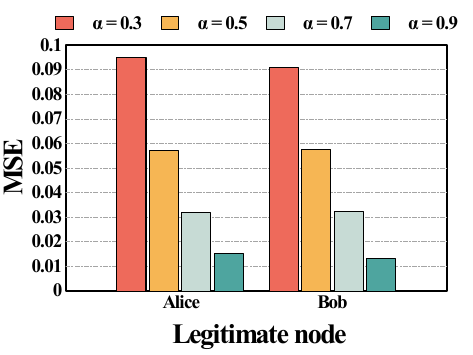}
\label{fig:eva_over2}
\end{minipage}%
}
\hspace{-0.1in}
\subfigure[Overall accuracy.]{
\begin{minipage}[t]{0.48\linewidth}
\centering
\includegraphics[width=1\linewidth]{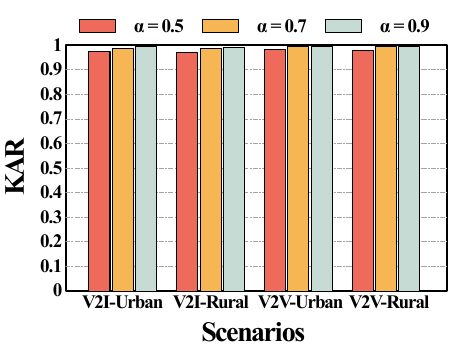}
\label{fig:eva_over}
\end{minipage}%
}
\vspace{-0.1in}
\centering
\caption{Overall Performance.}
\vspace{-0.3in}
\end{figure}

Fig.~\ref{fig:eva_over2} illustrates the average recovery accuracy of Alice and Bob for values of $\alpha$ ranging from 0.3 to 0.9. The results demonstrate that our system maintains an MSE below 0.1, even when the probing rate $\alpha$ is as low as 0.3. Notably, when $\alpha$ exceeds 0.7, the MSE drops to as low as 0.03.
Fig.~\ref{fig:eva_over} presents the key agreement rates obtained across V2I-Urban, V2I-Rural, V2V-Urban, and V2V-Rural scenarios using the method implemented in \SystemName. The figure further illustrates the key agreement rates at varying probing ratios, represented by $\alpha$, for each scenario.
The results indicate that \SystemName consistently achieves high key agreement rates across different scenarios and probing ratios. Notably, the system attains an average key agreement rate of 0.978 at $\alpha = 0.5$, 0.9915 at $\alpha = 0.7$, and 0.994 at $\alpha = 0.9$.
Overall, the findings highlight the exceptional effectiveness and robustness of \SystemName in diverse IoV scenarios.

\section{Conclusion}
\label{sec:conclusion}
In this paper, we propose \SystemName, a physical-layer key generation system designed for Next-Gen IoV systems. \SystemName addresses key inefficiencies found in existing wireless key generation methods, such as excessive channel probing and ineffective quantization. To overcome these challenges, we introduce several novel techniques to significantly enhance the performance of the key generation process.  
First, we develop an LLM-based channel probing method that reduces the number of probing rounds while preserving essential channel characteristics, enabling efficient and secure probing. Additionally, \SystemName replaces traditional quantization with a PCS-based key delivery mechanism, boosting both robustness and security.  
Real-world evaluations demonstrate that \SystemName achieves an average key agreement rate of 98.78\%.


\section*{Acknowledgment}
The project was supported by the NSF of Guangdong Province (Project No. 2024A1515010192)

\bibliographystyle{IEEEtran}
\bibliography{bib}

\end{document}